\documentstyle[12pt,boxedminipage,epsfig,wrapfig,floatflt,shadow]{article}
\begin{document}

\title{When physical intuition fails}
\author{Chandralekha Singh, University of Pittsburgh}
\date{ }

\maketitle

\vspace*{-.2in}

\begin{abstract}

We analyze the problem solving strategies of physics professors in a
case where their physical intuition fails. A non-intuitive
introductory-level
problem  was identified and posed to twenty physics professors.   The
problem placed the professors in a situation often encountered by
students, and their response highlights the 
importance of intuition and experience in problem solving.  While
professors had difficulty in solving the problem under the time
constraint, they initially employed a systematic approach, e.g.,
visualizing the problem,  considering various conservation laws, and
examining limiting cases.   After finding that familiar techniques were
not fruitful, they made incorrect predictions based on one of two
equally important factors.  By contrast, other more familiar problems
that require the consideration of two important principles (e.g.,
conservation of both energy and momentum for a ballistic pendulum) are
quickly solved by the same professors.  The responses of students who
were given the same problem reflected no overarching strategies or
systematic approaches, and a much wider variety of incorrect responses
were given.  This investigation highlights the importance of teaching
effective problem solving heuristics, and suggests that instructors
assess the difficulty of a problem from the perspective 
of beginning students.

\end{abstract}

\vspace*{-.4in}
\section{Introduction}
\vspace{-.2in}

Physical intuition is elusive--it is hard to define, cherished by those who possess it, difficult to convey to others.  
Physical intuition is at the same time an essential component of expertise in physics.  
Cognitive theory suggests that those with good intuition can efficiently search the information stored in memory to 
``pattern-match" or ``map" a given problem onto situations with which they have experience.
Over the course of their training, professional physicists develop a high degree of physical intuition that enables them to analyze and solve problems quickly and efficiently.  Standard introductory physics problems are ``easy" for professors because they know how to distill those physical situations into familiar canonical forms.  Introductory students often struggle over the same problems because they lack this ``distillation" ability, and because the canonical forms are not familiar.  

Problem solving can be defined as any purposeful activity where one is presented with a novel situation and devises and performs a sequence of steps to achieve a set goal~\cite{fred2}.  
The problem solver must make judicious decisions to reach the goal in a reasonable amount of time.  There is evidence to suggest that a crucial difference between the problem-solving capabilities of physics professors (experts)
and introductory physics students (novices) lies in both the level and complexity with which knowledge is represented and rules are applied~\cite{larkin,mestre,chi3}.  Physics professors view physical situations at a much more abstract level than beginning students, who often
focus on the surface features and get 
distracted by irrelevant details.  For example, students tend to group together all mechanics problems involving inclined planes, regardless of what type of physical principles are required for solving them~\cite{chi3}.  

Many studies have focused on investigating the differences between the
problem solving strategies used by expert physicists
and introductory physics students~\cite{larkin,mestre,chi3}. 
The problems chosen in these studies are typically those which physics professors find easy to solve using their intuition.
Here we analyze the problem solving strategies of physics professors in a case where their physical intuition fails.  
An introductory level problem was identified for which the physical intuition of most experts is lacking.  
We compare the problem-solving strategies of professors and introductory physics students in this context.
According to cognitive theory, expertise in a particular domain consists of having a large stock of compiled knowledge to deal with a wide variety of contingencies~\cite{anderson}.
No matter how expert people are at coping with familiar problems, their performance will begin to approximate 
that of novices once their stock of compiled rules in memory has been exhausted by the demands of a novel situation~\cite{anderson}. 
In these situations, experts cannot easily invoke 
compiled knowledge from memory because the applicability of a particular principle is not entirely obvious.
They must process information ``on the spot," in a manner similar to novices. 

We posed an introductory physics problem related to rotational and rolling motion 
to twenty physics professors and several introductory physics students. 
The question posed was inspired by a numerical problem found in the introductory textbook by Halliday, Resnick, and Walker~\cite{halliday}. 
It is interesting because despite being at an
introductory physics level, it is unlike the type of problems most professors have thought about before.
Of the twenty professors interviewed, not one had useful intuition that could guide them to the correct solution, 
nor could they easily identify how to solve the problem.

\vspace{-.3in}
\section{The problem on rotational and rolling motion}
\vspace*{-.2in}

\underline{Question :} Ignore the retarding effects of air resistance. A rigid wheel is spinning with an angular speed 
$\omega_0$ about a frictionless axis. The wheel drops on a horizontal floor, slips for some time and then rolls without slipping.
After the wheel starts rolling without slipping, the center of mass speed is $v_f$. How does $v_f$ depend upon the kinetic coefficient 
of friction $\mu$ between the floor and the wheel?\\
We suggest that the reader attempt to solve the problem before referring to the solution in Appendix  A.1 and A.2.

\vspace{-.3in}
\section{Details of the Study}
\vspace{-.2in}

The above problem was posed to twenty college professors.
Each discussion lasted between 15-30 minutes, depending upon the faculty member's interest in pursuing it. 
Although the discussions were not taped,  extensive notes were written down after each discussion to ensure that each faculty member's 
thought processes and problem solving strategies were captured accurately.
Another part of the study involves administering this problem~\cite{student} 
in the form of a recitation quiz to 67 calculus-based introductory physics students after they had taken an exam on rotational and rolling motion.
In addition to asking students to explain their reasoning, we discussed their intuition and approach individually with 
several students to better understand how they had interpreted and answered the problem.

\vspace{-.3in}
\subsection{Response of professors}
\vspace{-.2in}

After posing the problem, we asked the professors for their intuition along with their reasoning. 
If they were quiet for a long time, we encouraged them to articulate what they were thinking.
Most admitted that they did not have much intuition about how the final speed $v_f$ should depend on the coefficient of friction, $\mu$.  We
then asked them how they would approach the problem. Seven faculty went to the chalkboard and drew a picture of the situation.  Only three made an attempt to solve the problem quantitatively rather than reasoning qualitatively. This could be due to the fact that
they were asked about their intuition, and
they were hesitant in attempting a quantitative solution while under pressure since the direction/principle was not obvious.
While some were quick to point out their gut feelings and the corresponding reasoning, others were more cautious. 
Many noted that they did not have extensive experience dealing with problems in which the slipping part (rather than the rolling part) is
important. Three admitted having seen this type of problem before despite acknowledging a lack of intuition.
A few also mentioned that they were not good at { thinking} when put {``on the spot"}. 
Some expressed frustration at the fact that a simple conservation principle did not seem obvious for this problem.

What is fascinating about most professors' response is the manner in which they approached the problem. They almost always
visualized the problem globally and pondered over the applicable physics principles. More than half mentioned the idea of using some conservation principle,
{\it e.g.}, angular momentum conservation,
however, during the discussion, none could figure out how to apply it to the problem.
Many thought about the very high and low friction
limiting cases and several drew analogies with familiar situations which may employ similar underlying principles. 
Many invoked energy dissipation arguments.
However, seventeen out of the twenty professors concentrated almost exclusively on one of the two essential features of the problem, either the frictional force or the time to start rolling. 
The response of professors can be classified into five broad categories: 
(1) Five professors focussed on friction and noted that a larger friction would imply higher energy dissipation and, therefore, smaller $v_f$.
(2) Five professors focussed on the time to start rolling. They noted that a smaller friction would imply a larger slipping time before the wheel
locks. This would imply a larger energy dissipation leading to a smaller $v_f$.
(3) Three professors focussed on the fact that without friction
the wheel would keep slipping and never roll. Based upon this fact, they concluded that a larger friction implies a larger $v_f$.
(Although the conclusions in categories (2) and (3) are the same, we have separated them because professors in category (3) did not explicitly invoke slipping time or energy dissipation arguments.) 
(4) Three professors correctly observed that $v_f$ depends on two opposing factors: the time to start rolling and the the magnitude
of the frictional force. 
One of them believed that friction will dominate and a higher friction will imply smaller $v_f$ (he also noted that some conservation principle might be applicable, e.g., angular momentum conservation). Another
said that he was not sure which one of these opposing effects will dominate.
A third professor said that since a larger friction implies a larger acceleration but a smaller time before rolling, the {\it distance} travelled during slipping will be the same regardless of $\mu$. He suggested that a higher friction would probably imply a smaller $v_f$.
(5) Four did not express any clear opinion about whether $v_f$ should be larger or smaller if the frictional force is larger.
Three of them wondered whether the angular momentum conservation is applicable. However, they could not convince themselves about how and for which system this principle may be applicable.
Three of them considered the limiting cases (no friction implies that the wheel never rolls and very high friction implies it rolls immediately). 
Two briefly entertained all possible dependences of $v_f$ on $\mu$ but no clear reasoning was provided.

Professors often used reasoning that involved real-world analogies. One professor noted that the problem reminds him of airplane wheels during
landing. He said that he is wondering which principle is most appropriate in this case (for several minutes he made various hand 
gestures simulating the landing of a plane while trying to think about the applicable principle). 
He noted that the first thing that comes to his mind is the angular momentum conservation 
but then concluded that since the ground exerts a torque, the angular momentum cannot be conserved.  Then, he said that perhaps he should think 
about the energy dissipation but noted that it was not clear to him if the energy lost is higher when $\mu$ is higher or when the slipping
time is longer. 
Another professor who believed that a higher friction implies larger $v_f$ drew an analogy with walking. He said that while walking, the harder you
push the ground, the faster you can walk due to the reaction force of the ground. Similarly, the larger the frictional force that the
ground exerts on the wheel, the faster the $v_f$ should be. Immediately after being posed the problem, another professor drew an analogy with a 
pool ball which initially slips before rolling. He admitted that he did not have any intuition but drew a picture showing
the directions of $v$, $\omega$ and the frictional force and then wrote down the correct kinematic equations. 
He did not bother solving the equations but
said that since the acceleration is larger for the higher friction case while the time to start rolling is smaller, 
the distance travelled before rolling should be the same regardless of $\mu$ (an incorrect inference).

One professor recalled seeing this type of a problem in a textbook and noted that most likely angular momentum conservation should be used to 
identify the dependence of $v_f$ on $\mu$. However, during the discussions, he was unable to determine how to use this principle and 
for which system is the angular momentum conserved. Another professor who jokingly noted that he even remembers the page number on which
this problem is in a book said that he does not remember how $v_f$ should depend on $\mu$. He said that he was not sure whether the angular
momentum is conserved for this system and therefore he might attempt a kinematics route. He preferred
not to go to the chalk board during the discussion and said that he works best when not under pressure.
Another professor who preferred to go to the chalk board immediately drew the correct picture. He noted that no friction implies the wheel
never rolls while
high friction implies that it rolls immediately. He also noted that the frictional force $f_k$ increases $v$ since it is the only force on
the wheel and it decreases
$\omega$ since it causes a torque in a direction opposite to $\omega_0$. Then, he wrote down $f_k=\mu mg=ma$ amd $\tau =\mu mg  r= I \alpha$.
At this point, he tried to relate the linear and angular accelerations using $a=r \alpha$ (which is not correct since the wheel is not rolling
at this time). When it led to $I=mr^2$ which does not have to be true, he asked for more time to think about it. 
Another professor initially said that he is wondering whether there is a conservation principle, e.g., angular momentum conservation, that can be employed.
After pondering for sometime, he admitted that angular momentum conservation is often tricky to discern. 
Since he was not sure how to use it, he decided to use Newton's law/equation of kinematics but then he got confused about how to calculate
the linear acceleration of the wheel. 
He thought that friction should slow the wheel so there must be an additional force on the wheel that should increase its speed. 
He decided not to go to the chalk board.  Later, when we discussed the
problem solution, he admitted that drawing the picture would have helped. Pointing to the acceleration he jokingly said:
``this is where my intuition fails".

\vspace{-.3in}
\subsection{Student response}
\vspace{-.2in}

The student response can be classified in six broad categories: 
(1)  25 students ($37\%$) believed that friction will act in a direction opposite to the velocity and slow the wheel down. Therefore, larger $\mu$ 
implies smaller $v_f$. 
(2) 18 students ($27\%$) provided responses that were similar to the expert response category (2) and noted that since the frictional 
force is responsible for making the wheel roll, higher $\mu$ should imply higher $v_f$. 
(3) 6 students ($9\%$)
provided responses that were similar to the expert response category (3) and noted that since lower friction implies longer slipping time, the
$v_f$ will be lower in this case. (4) 4 students ($6\%$) provided reasoning different from that in category (1) to claim that higher $\mu$ would imply
smaller $v_f$. (5) 7 students ($10.5\%$) believed that $v_f$ will be independent of $\mu$ (which is the correct response) but only
one student provided qualitatively correct reasoning. 
(6) 7 students ($10.5\%$) provided responses which did not appropriately address the question that was asked.
For example, one noted that ``$v_f$ will be larger while the wheel is slipping and smaller when it grips."

Individual discussion shows that students seldom employed a systematic approach to problem solving and 
certain types of oversights common in student responses, were rare in the response of professors. 
Unlike professors, students rarely examined the limiting cases, contemplated the applicability of a conservation law or used analogical reasoning.
Many students did not take the time to visualize and analyze the situation qualitatively and they immediately 
jumped into the implementation of the solution based upon superficial clues. 
Many thought that the problem was relatively easy because there was friction on the floor and they were asked for the final speed of the wheel once
it starts rolling.
For example, $37\%$ of the students thought that friction will reduce the linear velocity because the two must oppose each other. 
Individual discussions show that several students in this category did not differentiate between linear and angular speed. When they were 
explicitly asked about whether there was a horizontal speed at the time the wheel hits the floor, some started to worry that they were confusing
the linear and angular speeds. 
Some assumed that the wheel 
will develop a linear speed as soon as it hits the ground.
When asked explicitly about what will cause it to develop the linear speed, some
noted that the impact will produce a linear speed as soon as the wheel hits the ground, others said
that there has to be a force in the direction of motion without actually identifying it, and a few admitted that they could not at the moment think of
a good reason for it. Inadequate time spent in visualizing the problem caused some students to confuse the vertical speed of the falling wheel with its horizontal speed.

Written responses and individual discussions show that many
students in {\it all} categories often focussed only on the linear speed and they largely ignored what changes the rotational speed to accomplish 
the rolling condition $v_f=r \omega_f$.
Such responses were rare from the professors. 
Professors almost always had a more holistic view of the problem, they always tried to visualize the problem,
and considered the changes in both the linear and angular speed to establish rolling.

\vspace{-.3in}
\section{Discussion}
\vspace{-.2in}

This investigation shows that even professors, who have a vast amount of physics knowledge, when forced to think ``on their feet" due to 
the novelty of the problem, have difficulties similar to those encountered by students in some ways.  
In solving problems about which they lack intuition, they have difficulty with the initial planning (decision making) of the problem solution. 
We emphasize that the problem posed was an introductory physics problem for which the planning of the solution only requires determining the
appropriate introductory physics concepts applicable in the situation. It does not involve invoking any techniques learned in
upper level or graduate courses.

The problem posed had two important variables that were inversely related to $v_f$: 
the force of friction and the time to start rolling.  Professors had great difficulty thinking about the effect of both parameters in the problem.  In particular, they often focussed only on one feature of the problem (friction or the time to roll) and did not consider the other one properly.
Those who focussed on the time to roll often noted that a high friction would lead to quicker rolling so less energy will be dissipated in that case and $v_f$ will be larger.  Those who focussed on friction and did not account for the time to roll, typically concluded that a high friction would lead to more energy dissipation and hence a smaller $v_f$.  Only three professors mentioned that both of the above factors will influence $v_f$.  Only one of them concluded that it was not obvious how $\mu$ will affect $v_f$. The other two ended up with incorrect inferences.

On the other hand, unlike students, professors in general had little difficulty considering the effect of friction on both the linear and rotational 
aspects of rolling motion simultaneously. Their training and experience made it quite natural to sense that
both types of motion will be affected by friction and consideration of both is important for establishing rolling.
The fact that in an unfamiliar situation, even professors struggled to focus on more than one important aspect of the problem while in a familiar
situation both aspects came naturally to them points to the importance of familiarity and experience in problem solving.

The rotational problem posed is analogous to one for which professors have no trouble intuiting the solution: the case of a completely inelastic collision between two objects.  In this case, the final speed is determined solely by linear momentum conservation, and is independent of the collision time.	To check the intuition of professors for the more familiar domain of linear motion, five of the twenty faculty members were asked about the completely inelastic collision of a bullet with a block resting on a horizontal surface.  They were asked about how the final speed of the bullet and the block moving together should depend upon the time it takes the bullet to come to rest with respect to the block due to the changes in the block material keeping its mass unchanged (if the material of the block is softer it will take longer for the bullet to come to rest with respect to the block). All of them responded correctly, noting that the linear momentum conservation guarantees that the time the bullet takes to come to rest with respect to the block is not relevant for determining the final speed of the block-bullet system moving together.  The spontaneity of expert respose to this problem, along with their difficulty in grasping how it may be applicable to the first problem posed to them, suggests that experience and familiarity with a particular type of problem are still very important in the problem solving skills of professors. 

Although professors behaved as students in some aspects, the problem solving strategies employed by them 
were generally far superior.  In particular, they often started by visualizing and analysing the problem qualitatively and searching for a useful conservation 
principles before resorting to other routes. They were much more likely to draw analogies and map the unfamiliar problem onto a 
familiar one.
They often examined limiting cases; a strategy which was rarely employed by students.
It is true that this problem excluded the zero friction limit because for that case the time for the wheel to start rolling is infinite. 
Thus, the final rolling condition is never met in this limit and the problem does not have a solution.
Therefore, relying on this limit does not yield useful clues and can lead to incorrect inferences as noted in several professors' response.
Nevertheless, examining the limiting cases and applicability of general principles is 
an excellent problem solving heuristic which can often make further analysis of the problem easier.
Some professors also mentioned or attempted to use kinematic methods. 
Despite their inability to solve the problem under time pressure, their holistic view and systematic problem solving approach
and knowledge-base helped them narrow down the problem space and prevented a wide range of oversights that were common in the student response.  
It was clear that while the their initial intuition was wrong, 
given enough time, their systematic approaches would invariably lead to the correct solution. 
On the other hand, a majority of students did not employ a systematic approach to problem solving. Individual discussions show that many students
jumped into the implementation of the solution without even taking the time to visualize the problem.
Several students thought that the problem was relatively straightforward because they only focussed on the fact that the effect of
friction on a final speed was required.
Many only focussed on the linear motion and they ignored what was responsible for changing the rotational motion to 
establish the rolling condition.
Professors adopted a much more global approach to the problem, and considered both the linear and rotational aspects of the problem.

The surprised reaction of several professors after finding out that $v_f$ is independent of $\mu$
hints at why the idealized situations, {\it e.g.}, motion on a frictionless surface,
are very difficult for students to internalize.
For example, one professor noted that he finds it very 
counter-intuitive because it implies that if the wheel fell on a low friction
ice and eventually started to roll, it will have the same $v_f$ as if it fell on a high friction surface. 
Of course, in a realistic situation, factors such as air-resistance and rolling friction will make $v_f$ 
dependent on $\mu$.
Only after one has carefully considered the limitations of
the idealizations in the light of our everyday experience can one feel comfortable making the corresponding inferences.

\vspace{-.3in}
\section{Summary}
\vspace{-.2in}

Expertise in physics is founded upon the pillars of intuition, knowledge, and experience.  Physicists continually transform their experiences into knowledge.  Intuition plays the role of a catalyst, greatly speeding up the process by allowing for shortcuts to be taken during problem solving.  
We have identified an introductory-level physics problem for which a group of twenty physics professors displayed a nearly universal lack of intuition.  
Although professors would have performed better without the time constraint,
our goal here was to elicit the thought-processes and problem solving strategies of experts as they venture into
solving a non-intuitive problem. In quizzes and examinations, students often work under a similar time constraint. 

The inherent difficulty of the problem posed in this study is comparable to problems professors can solve without
much difficulty.
This study suggests that 
the perceived complexity of a problem not only depends on its inherent complexity but also on the experience, familiarity, and intuition 
we have built about a certain class of problems. 
It has often been said that problems are either impossible or trivial, depending on one's success at solving them. 
Introductory students lack the vast experience, knowledge-base, and intuition that
the professors have about a majority of introductory physics problems.
As instructors, we should not be surprised 
that beginning students have great difficulty solving the ballistic pendulum problem, which 
requires invoking both the momentum and energy conservation principles. 
For professors, who have built an intuition about this class of problems, it appears
``easy"; for students, who lack intuition about these problems, it is difficult to focus on several aspects of the problem simultaneously.
There are likely to be less surprises if we put ourselves in students' shoes and analyze the difficulty of a problem from their perspective.

There are indeed few introductory level problems for which expert intuition is so universally lacking.  
The collective response of twenty professors to the problem suggests that none have frequently encountered or carefully thought about a 
problem like it.  A survey of most of the contemporary introductory textbooks supports this hypothesis.   
The response of {\it professors} in this study can shed some light on the kinds of difficulties that 
able students face as they solve problems and strive to develop physical intuition of their own.
Finally, although professors and students both had difficulties in solving the problem,
expert problem solving strategies were generally far superior. In particular, even 
when their intuition failed, professors started with useful heuristics such as visualizing 
and analysing the problem qualitatively, examining the applicability of convervation 
principles and limiting cases. 
While professors did not immediately know how to solve the problem, 
they demonstrated that they know how to solve problems, and given enough time, their systematic 
approaches would have inevitably led to the correct solution. 
It may be useful to design instructional strategies that {\it explicitly} teach problem solving heuristics 
and help students build and employ intuition in physical problems as we help our students 
learn various physics concepts.

\vspace{-.3in}
\section{Acknowledgments}
\vspace{-.2in}

We are very grateful to all of the faculty with whom the problem was discussed for their time 
and to F. Reif, R. Glaser, H. Simon (late), A. Lesgold, R. Tate, P. Shepard, and J. Levy for useful discussions.
We thank F. Reif, R. Glaser, J. Levy, and R. Johnsen for a critical reading of the manuscript.
This work was supported in part by the National Science Foundation and Spencer Foundation.
\vspace{-.1in}

\begin{center}
{\large{\bf APPENDIX A.1}}
\end{center}
\vspace{-.2in}

\underline{Answer:} $v_f$ is independent of $\mu$.

The fact that $v_f$ is independent of $\mu$ suggests the applicability of a conservation principle.
The problem can be viewed as a rotational inelastic collision with the floor, analogous to a linear inelastic collision.
We can invoke the conservation of angular momentum principle about the axis through the point 
where the wheel initially touches the ground. 
The angular momentum of the wheel is constant about this axis 
(during the time the wheel slips, there is a kinetic frictional force, but since the line of action of this force
passes through the axis, it does not produce a torque about the axis). 
Let $m$, $r$, and $I$ be the mass, radius, and
moment of inertia of the wheel about its center of mass respectively. 
For simplicity, we will assume that the wheel can be approximated as a hoop so that $I=m r^2$. 
Let $\omega_0$ be the initial angular speed of the wheel about its center of mass, and $v_f$ and $\omega_f$ be the linear and angular speed about its center of mass  
respectively when it starts to roll (see Figure 1). 
The initial angular momentum before the wheel touches the ground is just due
to the spin and $L_0=I \omega_0=m r^2 \omega_0$. 
When the wheel is rolling, the angular momentum about the chosen axis has two contributions:
one due to the spin and the other due to the linear motion $\vec r_{cm} \times (m \vec  v_f)$ where $\vec r_{cm}$ is the displacement
of the center of mass of the wheel from the chosen axis. The magnitude of the latter contribution is $r m v_f$ (see Figure 2) so that:
\begin{eqnarray}
L_f=I \omega_f + r m v_f=(I +m r^2) \omega_f=2 mr^2 \omega_f
\end{eqnarray}
where the rolling condition $v_f=r \omega_f$ has been used. Using the fact that $L_0=L_f$, $\omega_f=\omega_0/2$ independent of $\mu$ (see Figure 3).

Another approach to this problem is to use the equations of linear and rotational kinematics and the condition for rolling.
Let $t=0$ be the time when the wheel drops on the floor and $t$ be the time during which it slips before starting to roll. 
If the wheel is spinning in the clockwise direction when it drops on the floor, 
the frictional force will act to the right and will increase
its linear velocity (to the right) with a constant acceleration $a=F_k/m=\mu g$ (where $g$ is the magnitude of the acceleration due to
gravity) from its initial value zero (only spinning). 
The initial angular velocity $\omega_0$ will decrease with a constant angular acceleration $\alpha=r F_k/I=\mu g/r$ because the frictional 
force at the rim of the wheel causes a counter-clockwise torque $rF_k$.
Using equations of kinematics:
\begin{eqnarray}
v_f&=&at=\mu g t\\
\omega_f&=&\omega_0-\alpha t=\omega_0- \mu gt/r
\end{eqnarray}
Since the wheel starts to roll (without slipping) at time $t$, $v_f=r \omega_f$. Plugging the values of $v_f$ and $\omega_f$ from the 
above equations and solving, we obtain $t=\omega_0 r/(2 \mu g)$. Plugging $t$ in the above equations, $v_f=\omega_0 r/2$ and 
$\omega_f=\omega_0/2$, independent of $\mu$.

The above result can also be verified by noting that the energy dissipated by friction during slipping is independent of $\mu$.
Using the work-kinetic energy theorem, $W_{F_k}=K_f-K_i$ where $W_{F_k}=W_{lin}-W_{rot}$, $K_f=K_{f,lin}+K_{f,rot}$ and 
$K_i=K_{i,lin}+K_{i,rot}$ are the total work done by friction, and the total final and initial kinetic energies respectively.
$W_{lin}$,  $W_{rot}$, $K_{lin}$ and $K_{rot}$ are the work done by friction for the linear and rotational motion, and the linear and 
rotational kinetic energies respectively.
\begin{eqnarray}
W_{lin}&=&F_k x =mg\mu (g\mu t^2/2)=m \omega_0^2 r^2/8\\
W_{rot}&=&\tau \theta =F_k r \theta =mg\mu r (\omega_0 t - g\mu t^2/(2r))=3 m \omega_0^2 r^2/8
\end{eqnarray}
where $x$ and $\theta$ are the linear and angular displacements of the wheel respectively during the time it slips and we have used 
the equations of linear and rotational kinematics to relate $x$ and $\theta$ to $t$. 
Thus, the total energy dissipated by friction during slipping, $W_{F_k}= -m \omega_0^2 r^2/4$, is the same regardless of $\mu$ 
although the power (energy dissipated per unit time) depends upon it. For large $\mu$, the slipping time $t$ before the wheel starts to
roll is small but the power is high. This ensures that the total energy dissipated is the same in this case compared to that
when $\mu$ is small (and the slipping time is large). Therefore, $v_f$ is the same in both cases. 

\begin{center}
{\large{\bf APPENDIX A.2}}
\end{center}
\vspace{-.2in}

To show that $v_f$ is independent of $\mu$ regardless of the moment of inertia $I$ of the wheel,
we note that $\alpha=\mu m g r/I$ so that $\omega_f=\omega_0-\mu m g r t/I$. Using the condition for rolling without
slipping, $v_f=r \omega_f$, we obtain
\begin{eqnarray}
t&=&\frac{\omega_0 r}{\mu g (1+ mr^2/I)}
\end{eqnarray}
Plugging the value of $t$ in $v_f$ and $\omega_f$, we find that they are independent of $\mu$:
\begin{eqnarray}
v_f&=&\frac{\omega_0 r}{(1+ mr^2/I)}\\
\omega_f&=&\frac{\omega_0}{(1+ mr^2/I)}
\end{eqnarray}
We can calculate the total work done by friction and the work done for the linear and rotational components of motion with 
$x=a t^2/2$ and $\theta=\omega_0 t-\alpha t^2/2$ and find that they are independent of $\mu$:
\begin{eqnarray}
W_{lin}&=&F_k x =\frac{I^2 m r^2 \omega_0^2}{2(mr^2+I)^2}\\
W_{rot}&=&\tau \theta=\frac{I (mr^2)^2 \omega_0^2}{2(mr^2+I)^2} +\frac{I^2 m r^2 \omega_0^2}{(mr^2+I)^2}\\
W_{F_k}&=&W_{lin}-W_{rot}=- \frac{I\omega_0^2 mr^2}{2(mr^2+I)}
\end{eqnarray}
We can also calculate the change in the total kinetic energy of the system and show that it is equal to the total work done (independent of 
$\mu$):
\begin{eqnarray}
K_i&=& I \omega_0^2/2\\
K_f&=& (m v_f^2+ I \omega_f^2)/2=I \omega_0^2 \frac{I}{2(mr^2+I)}\\
W_{F_k} &=& K_f-K_i
\end{eqnarray}
The $I$ dependence (actually the dependence on the shape of the object because it is the ratio $I/(m r^2)$ that is important) of $\omega_f/\omega_0={1}/{(1+ mr^2/I)}$ is particularly interesting. In the limit as $I/(mr^2)-> 0$ (the mass of the object is localized close to the axis), $\omega_f -> 0$, so the maximal energy is dissipated by friction. The largest value $I$ can take is $I=mr^2$, which corresponds to the case in Appendix A. 
 Qualitatively, the dependence of $\omega_f$ on $I/(mr^2)$ can be understood by noting that less energy is dissipated if the angule speed has not decreased significantly when the rolling begins($v_f=r \omega_f$).
If the shape of the object is changed so that  $I/(mr^2)$ decreases while all other parameters are kept fixed, the angular speed will decrease more before the rolling condition is established.

The calculations can be repeated for the case where the initial linear speed is non-zero when the object touches the ground, 
{\it i.e.,}, $v_0 \ne 0$, but $\omega_0=0$ (as in the case of a non-spinning 
bowling ball thrown on the floor at an angle or a struck 
pool ball which initially only has a linear speed). 
The independence of $v_f$ on $\mu$ still holds (in fact, it holds even for cases where the object may initially have
both non-zero linear and angular speeds). Interestingly, in this case, 
the $I/(mr^2)$ dependence of $v_f$ and $\omega_f$ is opposite of the case noted in Appendices A.1 and A.2 ($\omega_0 \ne 0$ and  $v_0 =0$). 
Here, $v_f/v_0=1/(1+I/(mr^2))$. Therefore, in the limit as $I/(mr^2)->0$, 
$v_f=v_0$, so that negligible energy is dissipated by friction before the wheel starts rolling. 
Qualitatively, this can be understood by noting that less energy is
dissipated if the angular speed increases quickly to ``catch up" with the linear speed so that $v_f=r \omega_f$ without the 
linear speed having decreased significantly. Obviously, the angular speed will increase quickly if $I/(mr^2)$ is small.

\pagebreak

\begin{figure}
\begin{center}
\epsfig{file=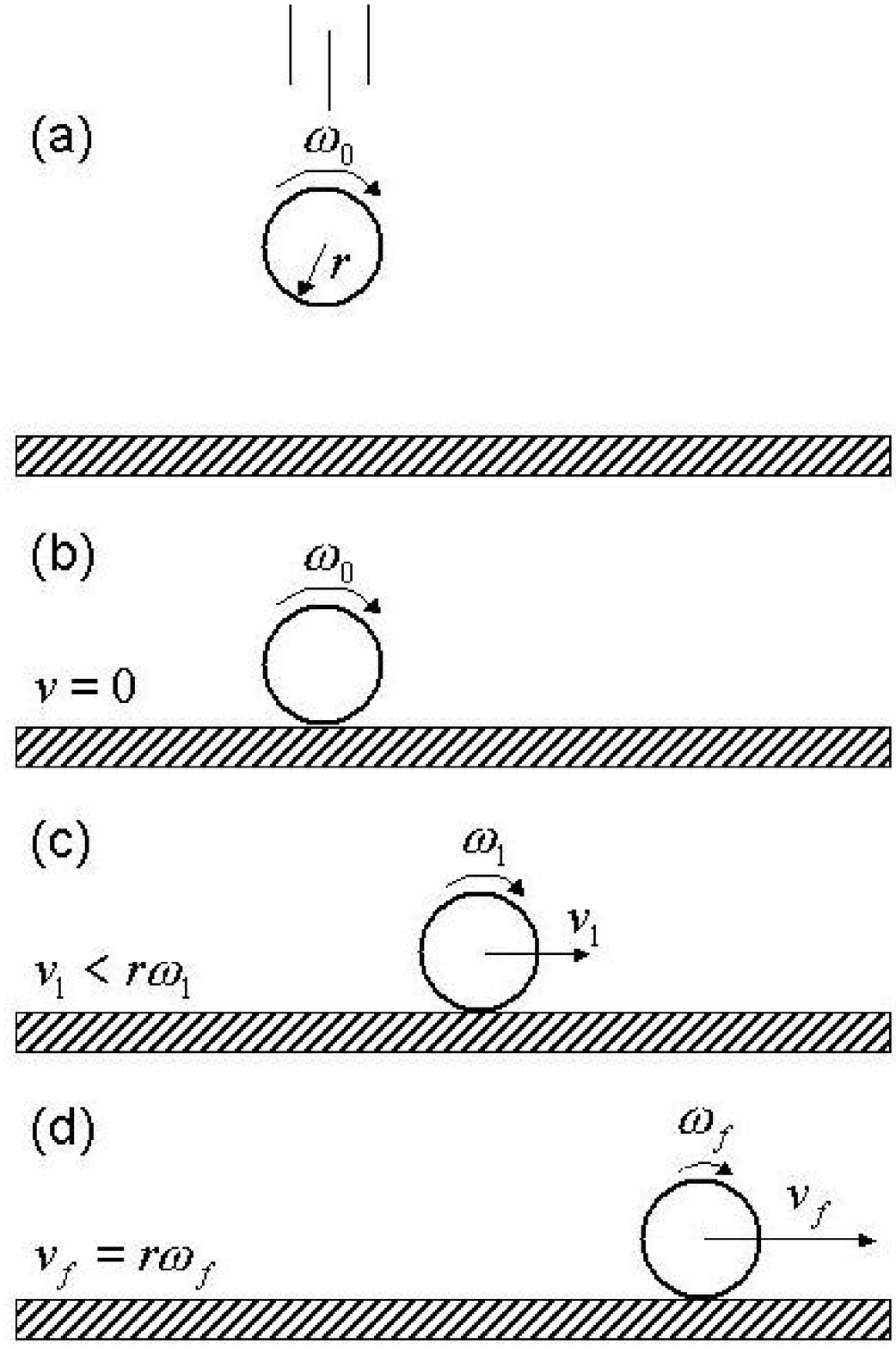,height=7.in}
\caption{Schematic diagram of the wheel at four different times: (a) spinning on a frictionless shaft, (b) 
hitting the floor, (c) slipping on the floor, and (d) rolling on the floor.}
\end{center}
\end{figure}


\begin{figure}
\begin{center}
\epsfig{file=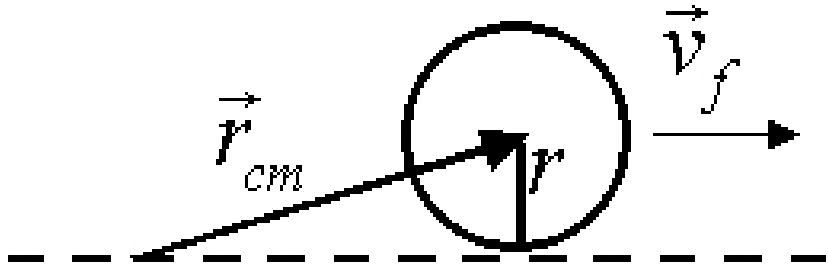,height=1.5in}
\caption{{Schematic diagram of the wheel showing $\vec r_{cm}$, $\vec v_f$ and $r$}. }
\end{center}
\end{figure}


\begin{figure}
\begin{center}
\epsfig{file=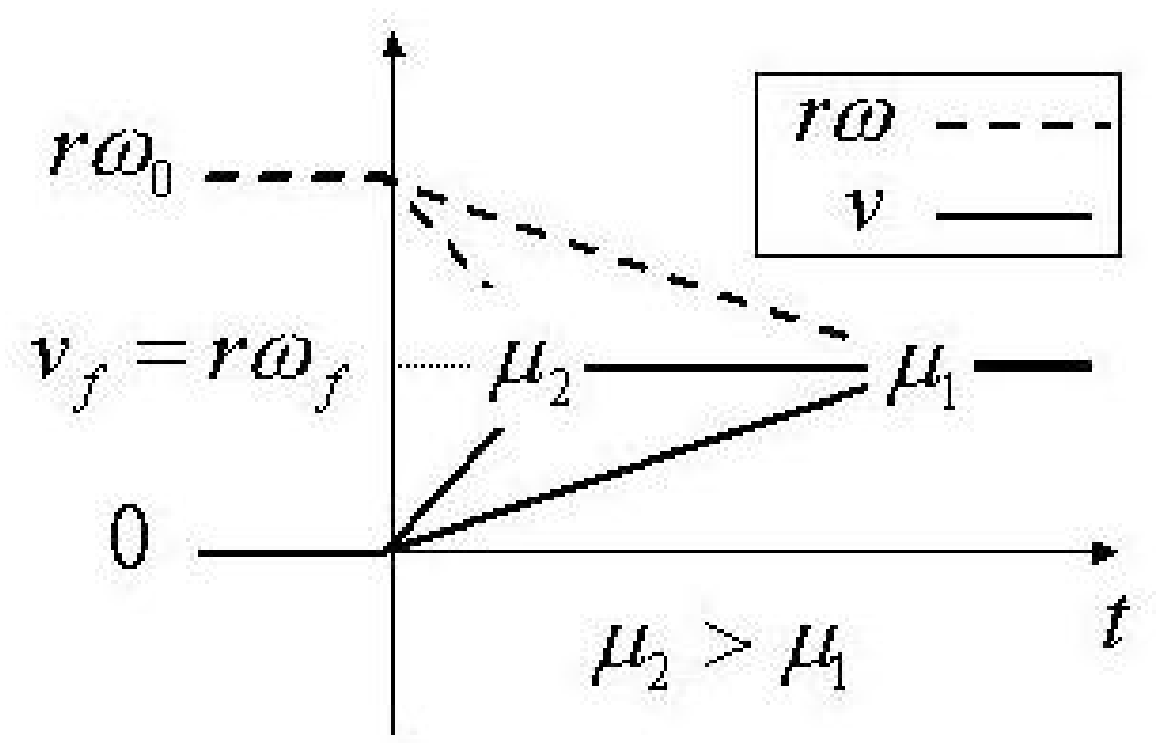,height=3.8in}
\caption{Graph of linear speed $v$ and scaled angular speed $r \omega$ versus time. Larger values of $\mu$ lead to shorter
locking times, but the final speed $v_f$ is independent of $\mu$.}
\end{center}
\end{figure}


\pagebreak


\begin{thebibliography}{9}
\bibitem{student} The problem posed to the students had the following additional sentences at the end: For example,
will $v_f$ be larger or smaller if the wheel fell on ice with small friction compared to a rough surface? Describe your intuition along with
the reasoning. 
You must draw a diagram and explain your approach to get full credit.
\end{thebibliography}
\end{document}